\documentstyle[aps,preprint,,multicol,epsf,rotate,citesort]{revtex}
\newcommand{\gl}[1]{Eq. (\ref{#1})}

\def\gtrless{\raise2.5pt\hbox{$>$}\llap{\lower2.5pt\hbox{$<$}}}
\def\gtrapprox{\raise2.5pt\hbox{$>$}\llap{\lower2.5pt\hbox{$\approx$}}}
\newcommand{\bsq}[1]{\begin{subequations}\label{#1}}
\newcommand{\esq}{\end{subequations}}
\newcommand{\beq}[1]{\begin{equation}\label{#1}}
\newcommand{\eeq}{\end{equation}}
\newcommand{\beqa}[1]{\begin{eqnarray}\label{#1}}
\newcommand{\eeqa}{\end{eqnarray}}

\newcommand{\vek}[1]{\vec{#1}}
\begin{document}

\title{Shear response of a smectic film stabilized by an external field}
\author{T.~Franosch and D.~R.~Nelson}
\address{Lyman Laboratory of Physics, Harvard University, Cambridge, Massachusets 02138}
\date{Draft, \today}
\maketitle

\begin{abstract}
The response of a field-stabilized 
two-dimensional smectic to shear stress is discussed. 
Below a critical temperature the smectic film exhibits elastic response 
to an infinitesimal shear stress normal to the layering. 
At finite stresses  free dislocations nucleate 
and relax the applied stress. The coupling of the dislocation current 
 to the stress
results  in non-newtonian viscous flow. 
The flow profile in a channel geometry is shown to change qualitatively 
from a power-law dependence to a Poiseuille-like profile  
opon increasing the pressure head.      
\bigskip

\medskip

\end{abstract}

%\begin{multicols}{2}

\section{Introduction}\label{Intro}

Thin layers of 
elongated molecules 
exhibit a variety of different 
liquid crystalline phases differing in the degree of positional  
and orientational order from isotropic films.
Complex rheological properties are common, 
in particular, when external orienting fields affect the alignment of 
the molecules. The coupling of the velocity field 
to the liquid-crystalline order gives rise to anisotropic viscosities. 
Furthermore, nonlinearities introduce shear-rate dependent effective
 viscosities, which includes phenomena like shear thinning, etc.

Recent experiments on Langmuir monolayers \cite{orient,Fuller,Schwartz} on 
an air-water interface
demonstrate the flow-induced 
microstructural deformations and orientation within a liquid-crystalline film. 
In particular, the non-linear shear response gives rise 
to unconventional flow profiles in channel geometry. 
Since dislocations often play a prominent role in two-dimensional systems, the 
motion of these point defects is presumably  responsible 
for many of the observed phenomena. 

Here we focus on a  smectic film floating on a liquid substrate. 
Smectic films are two-dimensional layered systems characterized by an 
 elastic free energy  
that descibes long-wavelength distortions of the layers in terms of 
 a displacement field $u(\vek{r})$ \cite{deGennes93}. Up to quadratic order,
we have
\begin{equation}    
\label{free-energy}
{\cal F} = \frac{1}{2} \int d^2 r [ B (\partial_z u)^2 + B \lambda^2 (\partial_x^2 u)^2 ] \, ,
\end{equation}
where the first term deals with layer compression, 
while the second term is related to splay distortions.  
The thermal fluctuations destroy even quasi-long-range translational 
order at any nonzero temperature. 
One easily finds that phonons give rise to exponential decay 
in the order parameter 
$\psi(\vek{r}) = \exp( i q_0 u(\vek{r}) )$, where $q_0 =2 \pi/d$ is 
the wavevector associated with the layer spacing $d$ \cite{Toner81}. In order 
to moderate the effect of thermal fluctuations one can 
apply an in-plane orienting external field. 
The corresponding free energy then has to be modified to \cite{Nelson83}
\begin{equation} 
\label{free}   
{\cal F} = \frac{1}{2} 
\int d^2 r [ B (\partial_z u)^2 + E (\partial_x u)^2 ] \, .
\end{equation}
Here $E$ measures the coupling to the external field, e.g. an 
electric or magnetic field. The nematic director orientation 
is aligned with the displacement field 
according to $\theta = - \mbox{const. }\partial_x u$ on large scales. This property does not 
necessarily hold near 
the boundaries. As discussed in the Appendix, there appears a 
characteristic length scale on which the director adjusts to the displacement.
   
One can eliminate the anisotropy in \gl{free} by a simple volume-preserving  
rescaling, and  the  static properties  
are thus determined by the universality class of the $XY$-model. 
There is a critical temperature $T_c \propto \sqrt{B E}$ separating a 
low-temperature smectic phase characterized by bound dislocations 
and power-law correlations in $ \langle \psi(\vek{r})^* \psi(0) \rangle$ 
and a high-temperature nematic phase with exponential decaying order parameter 
correlations  and free dislocations.   
Varying the external field allows control of the anisotropy. 
In particular, since the ratio $E/B$ is preserved 
under renormalization, one can adjust experimentally 
 the critical temperature $T_c$.
 
Here we discuss the dynamical properties of the stabilized 
smectic below $T_c$. The long-wavelength renormalized stiffness 
constants $E_R$ and $B_R$ now attain finite non-zero values and 
the film reacts as an elastic medium to an infinitesimal applied 
shear stress perpendicaular to the layers. However, a finite 
stress can trigger the proliferation of free 
dislocations which results in shear flow.    

\section{Hydrodynamics}
The long-wavelength and low-frequency dynamics of a smectic film is 
governed by hydrodynamic equations for the broken symmetry variable, viz. the 
layer displacement $u$, 
and the conserved quantities \cite{deGennes93}. For good thermal 
conductivity on a water substrate no temperature gradients can build up and 
energy conservation can be ignored. 
Furthermore we specialize to incompressible smectic films, 
setting the mass density $\rho = const$. Mass conservation then  
implies the constraint $\partial_x g_x + \partial_z g_z = 0 $ for 
the momentum density $\vek{g}(x,z)$. Similarly the trace of the viscous stress 
tensor also drops out of this long-wavelength low-frequency description.
The viscous stress tensor then has only two independent components
 which have to be related to the two of the strain rate tensor. 
From symmetry there appear two indpendent viscosities and we 
write the constitutive equation for the viscous stress as 
$\sigma_{xz}' = \nu (\partial_z g_x + \partial_x g_z), \sigma_{xx}' 
= (\nu-\nu') \partial_x g_x, \sigma_{zz}' = (\nu-\nu') \partial_z g_z$. 
The linearized equations of motion then read
\begin{mathletters}\label{disfree}
\begin{equation}
\label{ufield}
\partial_t u = g_z/ \rho + \lambda_p (B \partial_z^2 + E \partial_x^2 )u \, ,
\end{equation}
\begin{equation}
\partial_t g_x = - \partial_x p + \nu \partial_z^2 g_x + \nu'\partial_x \partial_z g_z \, ,
\end{equation}
\begin{equation}
\partial_t g_z = - \partial_z p + (B \partial_z^2 + E \partial_x^2 ) u + \nu \partial_x^2 g_z + \nu' \partial_z \partial_x g_x \, ,
\end{equation}
\end{mathletters}
where $\lambda_p$ denotes the permeation constant and $p$ the pressure.      

If a constant external shear stress $\sigma^{ext}_{xz}$ is applied, 
the medium reacts by building up a stationary 
viscous momentum flow $\sigma_{xz}^{ext} = \nu \partial_z g_x$. 
Since the layers are not distorted no elastic stresses can build up to balance the external stress. However, the medium does sustain  
 an infinitesimal 
external stress $\sigma^{ext}_{zx}$ by straining 
layers according to $\sigma_{zx}^{ext} = E \partial_x u$.      

The preceding paragraphs correctly describe the linear response of the 
stabilized smectic film. The nonlinear response, however, is 
qualitatively different. 
Finite applied stresses create free 
dislocations in the displacement field which can move and 
relax the stress. 
For two-dimensional smectics dislocations have Burgers' vector $\pm d$ 
along the $z$-axis (see Fig. 1). Hence, we can think of 
these dislocations as {\em scalar} defects with charge $m_\nu = \pm 1$. 
In the presence of plastic flow due to the dislocation motion 
the film 
now exhibits viscous behavior. A convenient way to incorporate 
the effects of dislocations is by switching from a description in 
terms of the displacement field $u$ to the strains $s_x, s_z$. 
This avoids the introduction of branch cuts in the displacement field 
$u(x,z)$. Locally the strains 
are given by $s_x = \partial_x u, s_z = \partial_z u$. However, since 
the displacement field is no longer  a single-valued function, the line 
integral $\oint_\Gamma (s_x dx + s_z dz ) $ does not vanish for closed 
loops $\Gamma$, but rather counts the number of  enclosed dislocations 
in units of the layer spacing $d$. Consequently the curl of the strain 
is given by $\partial_x s_z - \partial_z s_x = m(\vek{r}) d$, 
where $m(\vek{r}) = \sum_{\nu = \pm 1} m_\nu n_\nu(\vek{r})$
is the total ``charge'' 
dislocation density. The charges $m_\nu = \pm 1$  
 characterize the point dislocations and 
$n_\nu(\vek{r}) = \sum_{i_\nu} \delta(\vek{r}-\vek{r}_{i_\nu})$ is 
the number density of dislocations of charge $m_\nu$. Since dislocations can 
be created only in pairs, the overall dislocation density is conserved,
\begin{equation}
\label{conserve}
\partial_t m + {\rm div} \vek{J} = 0 \, .
\end{equation} 
Here $\vek{J}$ denotes the two-dimensional current of  dislocation charge. 
In terms of the strains, \gl{conserve},   
is merely the integrability condition which allows us to set 
$\partial_t s_z + J_x d = \partial_z \Xi, \, \partial_t s_x - J_z d = \partial_x \Xi$.
 Choosing $\Xi = g_z/\rho + \lambda_p (B \partial_z s_z + E \partial_x s_x)$ 
ensures compatability with the dislocation-free 
dynamics, \gl{ufield}. Upon collecting terms one derives
\begin{mathletters}\label{dis}
\begin{equation}
\label{sx}
\partial_t s_x  = \partial_x g_z/ \rho + \lambda_p \partial_x (B \partial_z s_z + E \partial_x s_x ) +J_z d\, ,
\end{equation}
\begin{equation}
\label{sz}
\partial_t s_z  = \partial_z g_z/ \rho + \lambda_p \partial_z (B \partial_z s_z + E \partial_x s_x ) - J_x d\, ,
\end{equation}
\begin{equation}
\label{gx}
\partial_t g_x = - \partial_x p + \nu \partial_z^2 g_x + \nu'\partial_x \partial_z g_z \, ,
\end{equation}
\begin{equation}
\label{gz}
\partial_t g_z = - \partial_z p + (B \partial_z s_z + E \partial_x s_x) + \nu \partial_x^2 g_z + \nu' \partial_z \partial_x g_x \, .
\end{equation}
\end{mathletters}

For the dislocation current we adopt a Fokker-Planck description. 
There is a diffusive current due to gradients in the dislocation density. 
Symmetry dictates that the principal axis of the diffusion tensor are 
aligned  with the $(x,z)$-coordinate system of the layers.  
Furthermore stresses induce motions that result in a separation 
of dislocations of opposite charge.  Following Ref. \cite{Toner81} we write
\begin{mathletters}\label{current}
\begin{equation}
\label{Jx}
J_x = \Gamma_x B d s_z n - \Gamma_x k_B T \partial_x m \, ,
\end{equation}
\begin{equation}
\label{Jz}
J_z = - \Gamma_z E d s_x n - \Gamma_z k_B T \partial_z m \, .
\end{equation}
\end{mathletters}
The first terms on the right-hand side 
are known as the Peach-Koehler forces and 
are the analog of the Magnus force in a superconductor, see Fig 1. 
The diffusion 
constants $D_x = \Gamma_x k_B T$ and $D_z = \Gamma_z k_B T $ refer 
to dislocation climb and glide. 
Since climb, i.e. motion perpendicular to the Burgers vector,
 involves long-range mass transport one 
expects $ \Gamma_x \ll \Gamma_z$. 
The free dislocation density 
$n(\vek{r}) = \sum_\nu m_\nu^2 n_\nu(\vek{r})$ vanishes 
in the equilibrium smectic phase 
since all dislocations are bound. Thus the {\em linear} 
dynamics given by Eqs. (\ref{dis}) and (\ref{current}) merely adds an 
anisotropic diffusive mode for dislocation motion 
to Eqs. (\ref{disfree}). 
%Since the displacement field corresponds 
%to a broken symmetry the weight of the diffusive line vanishes in the 
%long-wavelength limit. 

A non-zero free dislocation density $n(\vek{r})$ leads to 
viscous response to an external shear $\sigma_{zx}^{ext}$ as shown 
in Ref. \cite{Franosch2000}. Shearing perpendicularly to 
the layers of the smectic film to produce a nonzero 
$s_x \approx \partial_x u$  
liberates free dislocations which then can relax the shear. Thus 
one has a most prominent form of the phenomenon known as shear thinning, i.e. a reduction from an infinite linear viscosity and a non-zero linear  
elastic modulus to a finite nonlinear viscosity and zero nonlinear 
elastic shear modulus.  
This is the subject of the next two sections where we adopt  methods 
applied to superfluid helium films in 
 Ref. \cite{Ambegaokar80} to smectic hydrodynamics.

\section{Nucleation Rate of Free Dislocations}
The free energy of \gl{free} contains contributions from the smooth part 
of the strain and from the interaction energy of dislocations. 
The strains can be decomposed as 
$s_x = \partial_x \phi + s_x^{sing}, \, s_z = \partial_z \phi + s_z ^{sing}$. 
Then the smooth part and the dislocation interaction decouple provided 
one imposes the stress equlibrium condition 
\begin{equation}
E \partial_x s_x^{sing} + B \partial_z s_z^{sing} = 0 \, .
\end{equation}
This relation is in fact the integrability condition 
which allows the introduction of  the  Airy stress function 
$E s_x^{sing} = -\partial_z \chi, \, B s_z^{sing} =  \partial_x \chi$. 
Since the curl of the strain  yields the dislocation density, the  
Airy stress function is given in terms of the solution of 
\begin{equation}  
\label{Laplace}
\frac{1}{B} \partial_x^2 \chi 
+ \frac{1}{E} \partial_z^2 \chi = m(\vek{r}) d \, .
\end{equation}
After a partial integration the free energy, \gl{free}, can be written
\begin{equation}
{\cal F} = {\cal F}_0 - \frac{1}{2} \int d^2 r \chi(\vek{r}) m(\vek{r}) d \, ,
\end{equation} 
where the smooth fluctuations of the displacement field are encoded in 
\begin{equation}
{\cal F}_0 = \frac{1}{2} \int d^2 r [ B (\partial_z \phi)^2 + E (\partial_x \phi)^2] \, .
\end{equation}
The dislocation contribution is the smectic analog of the interaction energy of charges in electrostatics. Here, $m(\vek{r})$ corresponds to the electric 
charge density and  $\chi(\vek{r})$ to the electrostatic potential. 
As mentioned in Sec. \ref{Intro} the electrostatic analogy becomes complete 
after the volume prserving transformation 
$x = \alpha \xi, z =  \zeta/\alpha$, with $\alpha = (E/B)^{1/4}$. 
Equation (\ref{Laplace}) then reads
\begin{equation}  
 \partial_\xi^2 \chi 
+  \partial_\zeta^2 \chi = m  d B \alpha^2 \, .
\end{equation}
The bare 
interaction energy of a pair of dislocations of opposite charge $\pm 1$ 
is then easily calculated  to be 
\begin{equation}
\label{inter}
U_0(\varrho) = \frac{B \alpha^2 d^2}{2 \pi} \ln (\varrho/a_0) + 2 E_c(a_0) \, ,
\end{equation}
where $\varrho = \sqrt{\xi^2+ \zeta^2}$ is the distance of the dislocations 
in scaled units and $a_0 \sim d$ is a short distance cutoff that signals the 
breakdown of continuum elasticity theory. The core energy $2 E_c(a_0)$ 
plays the role of a chemical potential, i.e. the energy 
to create a defect 
pair at distance $a_0$ relative to a dislocation free system. 
Since there are many  pairs the effective interaction of a 
particular pair is a 
many-body problem. The pair carries a polarization cloud of bound pairs 
which screen the bare interaction. Following  Kosterlitz and Thouless 
\cite{Nelson83} 
this problem can be dealt with by the introduction of a scale-dependent 
dielectric constant $\epsilon(\varrho)$. The effective interaction then reads
\begin{equation}
\label{potential}
U(\varrho)  = \frac{B \alpha^2 d^2}{2 \pi} \int_{a_0}^\varrho \frac{d \varrho'}{\epsilon(\varrho') \varrho' }  + 2 E_c(a_0) \, .
\end{equation}     
The probability $\Upsilon(\varrho)$ per unit area 
to find a pair at distance $\varrho$ is then determined by the 
effective interaction 
$\Upsilon(\varrho) = a_0^{-4} \exp (- U(\varrho)/k_B T)$, or equivalently
\begin{equation}
\frac{d \Upsilon}{d \varrho} = - \frac{ B \alpha^2 d^2}{2 \pi k_B T \epsilon(\varrho) \varrho} \Upsilon(\varrho) \, .
\end{equation}
In terms of a scale  dependent stiffness 
$K(\varrho) = B \alpha^2 d^2 / [4 \pi^2 k_B T \epsilon(\varrho) ]$  
and the dislocation fugacity 
$y$, where $y(\varrho)^2 = \varrho^4 \Upsilon(\varrho)$, 
this is the first of the 
celebrated Kosterlitz recursion relations
\begin{equation}
\label{KT1}
\frac{d y}{d \ln \varrho} = [2 - \pi K ] y \, .
\end{equation}  
The renormalization of the stiffness is due to polarization of the dislocation pairs and is governed by the second recursion relation
\begin{equation}
\label{KT2}
\frac{d K^{-1}}{d \ln \varrho} = 4 \pi^3 y^2 \, .
\end{equation}
The flow equations for the dielectric constant and the fugacity
 then reveal the existence of a 
low-temperature phase with a finite long-wavelength dielectric constant.
Since $\epsilon(\varrho)$ depends intrinsically only on the flow parameter 
$l = \ln(\varrho/a_0)$  the first term in the 
effective potential is basically logarithmic and binds the pair  only weakly.

In the presence of a  strain $s_x$ the  pair is subjected to
the Peach-Koehler force $ E s_x  d$   that separates the pair 
in the $z$-direction in addition to the attraction force corresponding 
to \gl{potential}. 
For non-zero $s_x$ the total 
potential exhibits a saddle point on the $z$-axis and the pair can break 
up by escaping over the barrier as illustrated in Fig 2. 
This creates a pair of free 
dislocations that contributes to the relaxation of the applied shear.  

The motion of  bound pairs  is again given in terms of a 
Fokker-Planck equation \cite{Ambegaokar80}. 
Since climb motion is much slower than glide $\Gamma_x \ll \Gamma_z$ 
the associated 
currents are predominately unidirectional. With this simplification 
one derives
\begin{eqnarray}
\label{Fokker}
\partial_t \Upsilon & = & \partial_z {\cal J}_z \, , \nonumber \\
{\cal J}_z & = & 
- 2 \Gamma_z \Upsilon [\partial_z U +E s_x d]- 2 k_B T \Gamma_z  \partial_z \Upsilon  \, .
\end{eqnarray}
Since we are dealing with the {\em relative} motion of a pair, 
the  diffusion constant is $2 D_z$, where $D_z = \Gamma_z k_B T$ refers to 
diffusion of a
single dislocation. 
The escape rate of this effectively one-dimensional 
problem can be  obtained by 
standard methods. The saddle point $z_0 =  \zeta_0/\alpha $  
of the total potential   
determined by the implicit relation 
$ d  +   2 \pi \epsilon(|\zeta_0|)    \zeta_0 s_x \alpha = 0 $. 
For definiteness we discuss the case $s_x < 0$, i.e. 
the saddle point lies on the positive $z$-axis. 
The Fokker-Planck equation (\ref{Fokker}) implies 
that ${\cal J}_z$ is independent of the $z$-coordinate. The solution 
can be obtained in terms of $\psi = [U + E s_x z d] /k_B T $ by writing
\begin{equation}
\label{FP}
{\cal J}_z \int_z^\infty e^\psi dz' = -2 k_B T \Gamma_z e^\psi \Upsilon |_z^\infty  \, .
\end{equation}
For $z=+ \infty$ we expect that the dislocation probability 
density $\Upsilon$ vanishes rapidly, 
whereas for $|z| \ll z_0$ it assumes its equlibrium value. 
The right-hand side of \gl{FP} then simplifies to $2 k_B T \Gamma_z a_0^{-4}$. 
The left-hand side is dominated by the saddle point and the 
integral can be extended to $-\infty$. Upon expanding near the 
saddle point and using  the saddle-point equation, the 
Boltzmann weight can be approximated by 
\begin{equation}
e^\psi = \frac{\zeta_0^4 e^{-2 \pi K(\zeta_0)} }{ a_0^{4}  y(\zeta_0)^{2}} 
\exp[ \pi K(\zeta_0) \frac{x^2}{ \alpha^2 \zeta_0^2} -\pi K(\zeta_0) \frac{\alpha^2 (z-z_0)^2}{  \zeta_0^2}] \, .
\end{equation}
Here we neglected terms involving the derivative of 
the dielectric constant, since these 
are small on large scales according to \gl{KT2}.  
The dislocation current is then found 
as ${\cal J}_z = R \sqrt{K(\zeta_0)}/(\alpha \zeta_0) \exp[- \pi 
K(\zeta_0) x^2/(\alpha^2 \zeta_0^2)]$, where the production rate
$R = \int_{-\infty}^{\infty} {\cal J}_z dx $ reads
\begin{equation}
R = 2 k_B T \Gamma_z \alpha^2 y(\zeta_0)^2 \zeta_0^{-4} e^{2 \pi K(\zeta_0)} \,.
\end{equation}
For small strain $s_x$ and not too close to the critical temperature, 
one finds $\zeta_0 \gg \xi_-$, where $\xi_-$ is the 
correlation length implied by the  
recursion relations, Eqs. (\ref{KT1},\ref{KT2}). 
Then $K(\zeta_0)$ can be safely replaced by its 
large distance limit $K(\varrho = \infty) = 2/\pi [ 1 + x(T)/4]$. Here 
$x(T)$
measures the distance to the critical point, and for temperatures 
close to $T_c$ one has $x(T) \sim (T_c-T)^{1/2}$. 
Equation (\ref{KT1}) then implies in this regime 
$y(\varrho) \sim \varrho^{-x(T)/2}$ and correspondingly the production rate exhibits a power-law dependence on the strain
\begin{equation}
R \sim |s_x|^{4+x(T)} \, .
\end{equation} 

The dynamics of the free dislocation density is governed by the rate equation
\begin{equation}
\partial_t n = R - r n^2 \, .
\end{equation}
The recombination process of two free dislocations of opposite charge is 
encoded in the rate constant $r$. Since local equlibrium is established 
much faster than dynamics of the broken and conserved variables, one can assume that the free dislocation density follows the rate adiabatically, i.e, $n \sim R^{1/2} \sim |s_x|^{2+x(T)/2}$.

A stationary, uniform external stress $\sigma_{zx}^{ext}$ is 
balanced by a uniform strain $ E s_x$, \gl{gz}. This results in a 
nonlinear dislocation flow $J_z \sim s_x^\gamma, \, 
\gamma = {3+x(T)/2}$, which gives rise 
to shear strain rates
 $\partial_x g_z = - J_z \rho d$. Upon collecting 
terms one derives the nonlinear constitutive equation
\begin{mathletters}
\begin{equation}
\partial_x g_z \sim (\sigma_{zx}^{ext})^{\gamma} \, ,
\end{equation} 
\begin{equation}
\gamma = 3 + x(T)/2 \, .
\end{equation}
\end{mathletters}

\section{Channel flow}
In this section we discuss the application of these ideas
to flow of the stabilized smectic due to a
 pressure gradient along a channel. The film layers are oriented 
perpendicularly to the channel flow, i.e. the pressure gradient is 
along the $z$-direction.
Furthermore we assume that the channel is much wider than 
the correlation length $\xi_-$ so that the hydrodynamic description is valid. 

In the steady state all $z$-derivatives vanish due to translational 
invariance along the channel, except for the 
pressure head $\pi' = - \partial_z p = const$. The equations of motion,
 Eqs. (\ref{dis}), then reduce to 
\begin{mathletters}\label{channel}
\begin{equation}
0 = \partial_x g_z/ \rho + \lambda_p   E \partial_x^2 s_x  +J_z d\, ,
\end{equation}
\begin{equation}
0 = \pi' +  E \partial_x s_x + \nu \partial_x^2 g_z  \, .
\end{equation}
\end{mathletters}
Furthermore we impose the no-slip boundary
 condition $g_z( \pm a) = 0$ and 
vanishing permeation current $\partial_x^2 s_x( \pm a) = 0$, 
where $\pm a$ are the walls of the channel. Symmetry then 
implies $\partial_x g_z(0)= \partial_x^2 s_x(0) =s_x(0) = 0$. 
The derivative  $\partial_x g_z$ can now 
be eliminated from \gl{channel} yielding
\begin{equation}
- \pi' x/ E = s_x - \delta^2 \partial_x^2 s_x- \rho \nu J_z d/E \, , 
\end{equation}
where we introduced the 
permeation length $\delta = (\rho \nu \lambda_p)^{1/2}$. 
Since the dislocation current is nonlinear in the strain, the flow 
profile depends on the pressure head. 
The  combination $\rho \nu J_z d/E $ can be written as $-  s_x |A s_x|^{\gamma-1}$, where $A$ is a dimensionless constant. Rescaling of $s_x$ reveals another characteristic length scale $L_\pi = E/(A \pi') $, apart from the channel width and the permeation length. 
 There are several 
cases that can be discussed analytically.  

For  $L_\pi \gg a $ 
the  pressure head is mostly balanced by 
elastic deformations of the smectic and 
the strain is given approximately by $s_x = - \pi' x/E$. 
The dislocation current is negligible compared to the elastic 
contributions, however, it gives rise to the fluid flow $\partial_x g_z = E 
s_x | A s_x|^{\gamma-1}/\nu$. The flow profile  then has the ineresting 
non-Newtonian form
\begin{equation}
g_z(x) = \frac{(\pi')^\gamma  A^{\gamma-1}}{(\gamma+1) \nu E^{\gamma-1}} [ a^{\gamma+1}- |x|^{\gamma+1} ] \, ,
\end{equation}    
where $\gamma$ is a continuously variable exponent in the range 
$0 \leq \gamma = 3+x(T)/2 \leq 3$.
In particular, the velocity is non-linear in the pressure head 
contrary to conventional Poiseuille flow.

In the opposite case $L_\pi \ll a$, we assume 
that the permeation current can be neglected, except for a boundary layer of order $\delta$. 
Then for $|x| \ll L_\pi$, one finds again $s_x = - \pi x/E$ whereas 
for $|x| \gg L_\pi$ one observes 
Poiseuille behavior $\partial_x g_z = - \pi' x/\nu$. The solution is 
self-consistent provided $\delta \ll a (a/L_\pi)^{(\gamma-1)/(2\gamma)}$.

\section{Conclusion}

We derived the hydrodynamic equations of motion for a smctic film stabilized by an external field. Below the Kosterlitz-Thouless transition 
dislocations are bound in pairs. However, a finite shear stress applied perpendicularly to the smectic layers  
nucleates dislocation pairs. 
The motion of these  free dislocations results 
in a non-linear viscous response. The smectic film exhibits 
shear-thinning giving rise to unusual flow profiles in channel flow.

\acknowledgments{
We would like to acknowledge conversations with S. Jain during the 
early stages of this investigation. 
This research was supported by the National Science Foundation, through 
the MRSEC Program via
Grant DMR-98-09363 and through Grant DMR-9714725. T.F. acknowledges the 
support   by  the Deutsche
Forschungsgemeinschaft under Grant No.  Fr 417/2.}

\appendix
\section{Alignment of the nematic director with the displacement field}
 
The free energy \gl{free} is the long-wavelength description of the 
smectic film. The nematic director $\theta$ is locked to the 
displacement via $\theta = - \mbox{const. } \partial_x u$. This 
can be seen by taking a 
more microscopic approach by a free energy that includes the nematic director
degrees of freedom explicitly, \cite{Toner81,Nelson83}
\begin{equation}
\label{freemicro}
\tilde{\cal F} = \frac{1}{2} \int d^2 r [ B(\partial_z u)^2 + D (\partial_x u + \theta)^2 + K_0 (\vek{\nabla} \theta)^2 + \tilde{E} \theta^2 ] \, .
\end{equation}  
Here $D$ measures the free energy cost for nonaligned director to 
displacement, $K_0$ is the Franck constant and $\tilde{E}$ the coupling of 
the director to the external field. For an infinite system the contribution 
the director field can be integrated out leading to an effective free energy
in terms of the Fourier transformed displacement field 
\begin{equation}
{\cal F} = \frac{1}{2} \int_q [ B q_z^2 + D q_x^2 - \frac{D^2 q_x^2}{\tilde{E} + D q_x^2 + K_0 q^2}] |u(\vek{q})|^2 \, .
\end{equation}   
Expanding to lowest nontrivial order in powers of the wavevector $\vek{q}$ 
leads to the free energy \gl{free} with $E = D \tilde{E}/(D+ \tilde{E})$.

Near the boundary we use the more microscopic free energy \gl{freemicro} and 
derive the corresponding Euler equations
\begin{mathletters}
\begin{equation}
D \partial_x u - K \nabla^2 \theta + (D +\tilde{E}) \theta = 0 \, .
\end{equation}
\begin{equation}
B \partial_z^2 u +D \partial_x^2 u - D\partial_x \theta =0 \, ,
\end{equation}
\end{mathletters}
The first equation reveals the characteristic length scale $\lambda_0
 = \sqrt{K/(D+\tilde{E})}$.  For 
variations of the displacement field with 
wavevectors $ q \ll \lambda_0^{-1}$ the director field is 
given by $\theta = - \partial_x u D/(D+\tilde{E})$. 
Near the boundary where independent boundary conditions on the director and 
the displacement field can be imposed is a layer of order $\lambda_0$ where 
the previous relation does not hold.  
The second equations shows that the  parameter entering the Euler 
equations corresponding to the effective free energy 
is $E = D \tilde{E}/(D+\tilde{E})$.

%\end{multicols}

\begin{figure}
\caption{Layering of the elongated molecules of a two-dimensional smectic. 
The  dislocation 
on the right half has positive ``charge'', whereas the dislocation 
on the left possesses negative ``charge''.  
The strain $s_x$ exerts a Peach-Koehler force on the free 
dislocations indicated by the arrows. 
The induced dislocation current  then relaxes the strain. }
\end{figure}

\begin{figure}
\caption{Sketch of the 
effective potential for a dislocation pair 
in the presence of shear $k_B T \psi = U+E s_x z d$. 
The heavy line marks the boundary of bound and free dislocation pairs.
Breaking of a bound pair occurs via escape over the saddle point. 
Since climb motion is negligible compared to 
glide the $x$-coordinate is effectivley frozen.  }
\end{figure}

\newpage

\pagestyle{empty}
\setcounter{figure}{0}

\begin{figure}
\centering
\leavevmode
\parbox{17cm}{

\epsfxsize=6.2in
\epsfysize=6.2in
\epsfbox{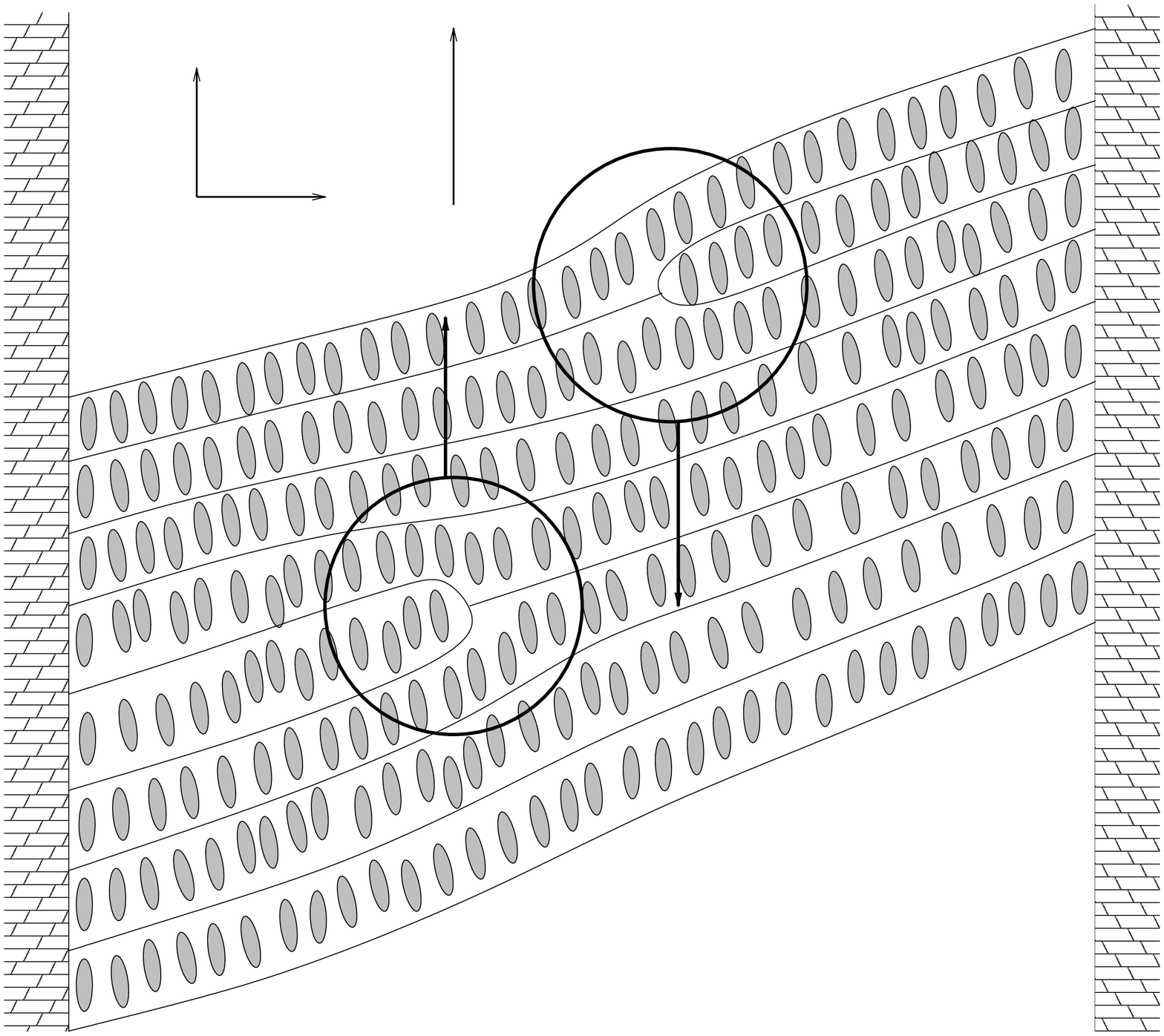}

\vspace{-14cm}\hspace{4.5cm}{\Large $x$}

\vspace{-3cm}\hspace{2.0cm}{\Large $z$}

\vspace{0.5cm}\hspace{6.8cm}{\Large $\vec{E}$}

\vspace{15cm}

\vfill
\caption{}
}
\end{figure}

\newpage

\pagestyle{empty}
\setcounter{figure}{1}

\begin{figure}
\centering
\leavevmode
\parbox{17cm}{

\epsfxsize=6.2in
\epsfysize=6.2in
\epsfbox{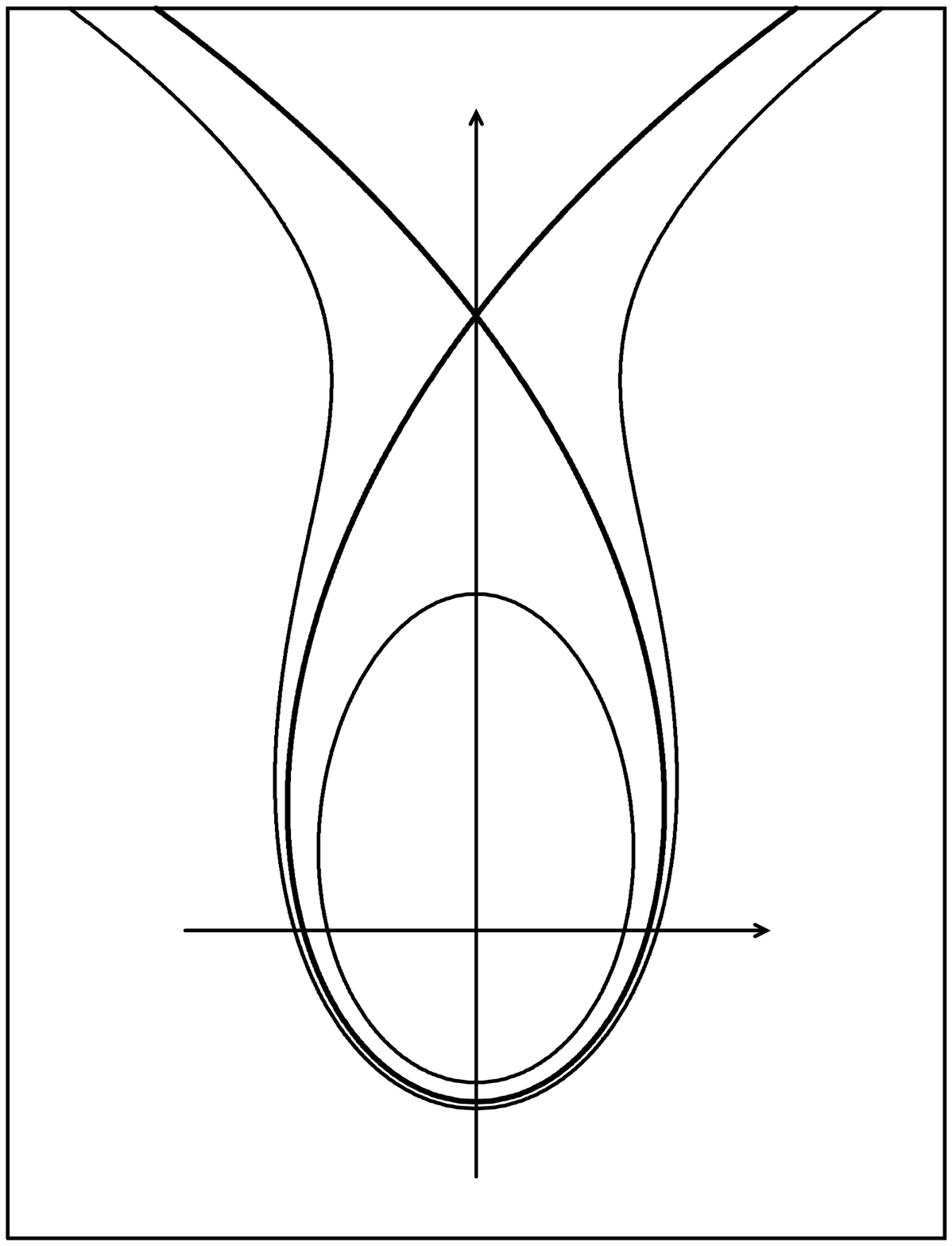}

\vspace{-5cm}\hspace{12.3cm}{\Large $x$}

\vspace{-10cm}\hspace{8.3cm}{\Large $z$}

\vspace{1.5cm}\hspace{7.0cm}{\Large $z_0$}

\vspace{1.5cm}\hspace{6.3cm}{\Large bound}

\vspace{1.5cm}\hspace{3.5cm}{\Large free}\hspace{7cm}{\Large free}

\vspace{7cm}

\vfill
\caption{}
}
\end{figure}


\begin{thebibliography}{10}

\bibitem{orient} C. Mingotaud, B. Agricole, and C. Jego, 
J. Phys. Chem. {\bf 99}, 17068 (1995).

\bibitem{Fuller} T. Maruyama, G. Fuller, C. Frank, and C. Robertson, 
Science {\bf 274}, 233 (1996).

\bibitem{Schwartz} M. L. Kurnaz and D. K. Schwartz, Phys. Rev. E {\bf 56}, 
%No. 3, 
3378 (1997).

\bibitem{deGennes93}
P.G. de Gennes and J. Prost, {\em The Physics 
of Liquid Crystals} (Oxford Univ. Press, Oxford, 1993), 2nd ed.



\bibitem{Toner81}
J. Toner and D.~R. Nelson, Phys. Rev. B {\bf 23},  316  (1981).


\bibitem{Nelson83}
See, e.g., D.~R. Nelson in {\em Phase Transitions and Critical Phenomena}, 
Vol. 7, edited by C. Domb and J. Lebowitz (Academic, London, 1983) pp. 76-79

\bibitem{Ambegaokar80}
V. Ambegaokar and B.~I. Halperin and D.~R. Nelson and E.~D. Siggia, Phys. Rev. B {\bf 21}, 1806 (1980)       

\bibitem{Franosch2000}
T. Franosch and S. Jain and D.~R. Nelson, Phys. Rev. E {\bf 61}, 3942 (2000)
\end{thebibliography}
\end{document}